\documentclass[showpacs,floatfix,prd,
twocolumn,superscriptaddress,nofootinbib]{revtex4}
\usepackage{graphicx,subfigure,bm,color,psfrag,hyperref}
\usepackage{amsfonts}
\usepackage{lipsum}
\usepackage{mathtools}
\usepackage{verbatim}
\usepackage[normalem]{ulem}
\usepackage[dvipsnames]{xcolor}
\usepackage{amsmath}
\usepackage{multirow}
\usepackage{soul}
\usepackage{dcolumn}
\usepackage{amssymb}
\usepackage{amsbsy}
\usepackage{rotating}
\usepackage{overpic}
\usepackage{ifxetex}
\usepackage{lipsum}
\usepackage{float}
\usepackage{bbold}
\usepackage{enumerate}
\usepackage{dsfont}
\usepackage{mathrsfs}
\usepackage{mathtools}
\usepackage{tensor}
\usepackage{caption}
\usepackage{subcaption}
\usepackage{ragged2e}

\hypersetup{colorlinks,linkcolor={blue},citecolor={red},urlcolor={teal}}  

\definecolor{amaranth}{rgb}{0.9, 0.17, 0.31}
\definecolor{palatinateblue}{rgb}{0.15, 0.23, 0.89}
\definecolor{brightpink}{rgb}{1.0, 0.0, 0.5}
\hypersetup{ linktoc=all,
    colorlinks, linkcolor={palatinateblue},
    citecolor={brightpink}, urlcolor={amaranth}
}
\graphicspath{ {images/} }

\newcommand{\be}{\begin{equation}}
\newcommand{\ee}{\end{equation}}
\newcommand{\ba}{\begin{eqnarray}}
\newcommand{\ea}{\end{eqnarray}}

\def\doi{http://doi.org}

\begin{document}

\title{Equivalence of $f(Q)$ cosmology with quintom-like scenario: 
the phantom field as effective realization of the non-trivial connection}

\author{Spyros Basilakos}
   \email{svasil@academyofathens.gr}
    \affiliation{National Observatory of Athens, Lofos Nymfon, 11852 Athens, 
Greece}
\affiliation{Academy of Athens, Research Center for Astronomy and Applied 
Mathematics, Soranou Efesiou 4, 11527, Athens, Greece}

\affiliation{School of Sciences, European University Cyprus, Diogenes 
Street, Engomi, 1516 Nicosia, Cyprus}

\author{Andronikos Paliathanasis}
\email{anpaliat@phys.uoa.gr}
\affiliation{Institute of Systems Science \& Department of Mathematics, Faculty 
of Applied Sciences, Durban University of Technology, Durban 4000, South Africa}
\affiliation{School for Data Science and Computational Thinking, Stellenbosch University,
44 Banghoek Rd, Stellenbosch 7600, South Africa}
\affiliation{Departamento de Matem\'{a}ticas, Universidad Cat\'{o}lica del 
Norte, Avda.
Angamos 0610, Casilla 1280 Antofagasta, Chile}

\author{Emmanuel N. Saridakis}
\email{msaridak@noa.gr}
\affiliation{National Observatory of Athens, Lofos Nymfon, 11852 Athens, Greece}
\affiliation{Departamento de Matem\'{a}ticas, Universidad Cat\'{o}lica del 
Norte, Avda.
Angamos 0610, Casilla 1280 Antofagasta, Chile}
\affiliation{CAS Key Laboratory for Researches in Galaxies and Cosmology, Department of
Astronomy, University of Science and Technology of China, Hefei, Anhui
230026, P.R. China}

\begin{abstract}

We show that $f(Q)$ cosmology with a non-trivial connection,
namely the Connection II of the literature, is dynamically equivalent with a
quintom-like model. In particular, we show that the scalar field arising
from the non-linear $f(Q)$ form, and the scalar field associated to the
non-trivial connection, are combined to provide one canonical and one phantom
field in the minisuperspace Lagrangian. Hence, the combination of two
well-behaved fields appears effectively as a phantom field. This is 
fundamentally different from the usual approach in phantom and quintom 
cosmology, in which the phantom field is introduced ad hoc, and thus it may 
open a
new way to handle the phantom fields, namely as effective realizations of
the connection-related canonical fields.

\end{abstract}

\maketitle

\renewcommand{\tocname}{Index}

\section{Introduction}

In an effort to address certain theoretical issues of General Relativity,
such as its non-renormalizability \cite{Stelle:1976gc}, as well as the
cosmological constant problem \cite{Weinberg:1988cp}, but also to provide
potential ways to alleviate various observational tensions \cite%
{Perivolaropoulos:2021jda,Abdalla:2022yfr}, a wide range of modified gravity
theories have been developed \cite{CANTATA:2021asi}. There exist various
avenues for constructing such modifications. A common approach begins with
the standard curvature-based formulation of gravity, leading to theories
such as $f(R)$ gravity \cite{Starobinsky:1980te,DeFelice:2010aj}, $f(G)$
gravity \cite{Nojiri:2005jg}, $f(P)$ gravity \cite{Erices:2019mkd}, and
Lovelock gravity \cite{Lovelock:1971yv}, among others. Alternatively, one
may consider the torsional formulation of gravity, which gives rise to
frameworks like $f(\mathbb{T})$ gravity \cite{Bengochea:2008gz,Cai:2015emx}
and $f(\mathbb{T},\mathbb{T}_{G})$ gravity \cite{Kofinas:2014owa}. However,
there is a third powerful path that has attracted the interest of the
community, and involves the use of non-metricity as the fundamental
geometric quantity \cite{Nester:1998mp,BeltranJimenez:2017tkd}, leading to
the construction of $f(Q)$ gravity theories \cite%
{BeltranJimenez:2019tme,Anagnostopoulos:2021ydo,Lazkoz:2019sjl,Lu:2019hra,
Mandal:2020buf,Ayuso:2020dcu, Frusciante:2021sio,
Ferreira:2022jcd,Gadbail:2022jco,
Sarmah:2023oum,Khyllep:2021pcu,Barros:2020bgg, De:2022jvo,Solanki:2022ccf,
DAmbrosio:2021zpm,Li:2021mdp,Dimakis:2021gby, 
Kar:2021juu,Wang:2021zaz, Quiros:2021eju,Mandal:2021bpd,Albuquerque:2022eac,
Anagnostopoulos:2022gej,Arora:2022mlo,Pati:2022dwl,
Dimakis:2022wkj,DAgostino:2022tdk,
Narawade:2022cgb,Emtsova:2022uij,Bahamonde:2022cmz,
Sokoliuk:2023ccw,Shaikh:2023tii,Dimakis:2023uib,
Koussour:2023rly,Najera:2023wcw,Atayde:2023aoj,
Shabani:2023xfn,Tayde:2023xbm, De:2023xua,Junior:2024xmm, 
Millano:2024rog,Yang:2024tkw,Guzman:2024cwa, 
Alwan:2024lng,Shaily:2024tmx,Moreira:2024unj,Dubey:2024gxa, 
Mushtaq:2025jjj,Nashed:2025usa, 
Papagiannopoulos:2025uix,Moreira:2025rxp,Dimakis:2025jrl,youri}
(for a review see \cite{Heisenberg:2023lru}).

On the other hand, the cosmological constant problem, as well as the
observational issues of the Standard Model of Cosmology, may be addressed
maintaining general relativity as the underlying gravitational theory but
introducing extra scalar fields, with the notable examples being the
inflaton \cite{Lidsey:1995np} and the quintessence \cite{Ratra:1987rm}.
Although the scalar fields are typically considered as canonical ones, the
need to describe the \textquotedblleft phantom\textquotedblright\ regime,
namely to obtain equation-of-state parameters smaller than the cosmological
constant value $-1$ led to the wide use of phantom fields too, namely fields
that have a negative kinetic energy in the Lagrangian \cite{Singh:2003vx}.
Despite the fact that the phantom fields can have many interesting cosmological
implications and can lead to viable phenomenology \cite%
{Dabrowski:2003jm,Chimento:2003qy,Sami:2003xv,Gonzalez-Diaz:2003bwh,
Caldwell:2003vq,Elizalde:2004mq,Perivolaropoulos:2004yr,Nojiri:2005pu,
Faraoni:2005gg,Nojiri:2005sx,Gannouji:2006jm,Chen:2008ft,Elizalde:2008yf,
Saridakis:2009pj,Astashenok:2012tv}, their introduction is ad hoc, and the
fact that their energy is unbounded from below makes them unstable at the
quantum level \cite{Schon:1981vd,Witten:1981mf,Cline:2003gs,Saridakis:2008fy}. 
However, although there have been efforts to construct a phantom theory 
consistent
with quantum field theory, for instance trying to make the phantom fields to
arise as effective descriptions of higher-dimensional canonical theories 
\cite{Nojiri:2003vn, Nojiri:2003ag}, the whole issue is still open. Finally,
the simultaneous consideration of one canonical and one phantom field, gives
rise to the two-field, quintom model, which shares the advantages of both
quintessence and phantom cosmology and moreover is able to describe the
phantom-divide crossing \cite
{Cai:2009zp,Feng:2004ad,Guo:2004fq,Zhao:2005vj,Zhao:2006mp,Cai:2006dm,
Guo:2006pc, Cai:2007qw}. We mention here that the phantom-divide crossing is 
strongly favoured by various datasets, among others by the recently released 
DESI DR2 data \cite{Lodha:2025qbg}.

In this work we show that $f(Q)$ cosmology with a non-trivial connection,
namely the Connection II of the literature, is dynamically equivalent with a
quintom-like model. In particular, we show that the scalar field arising
from the non-linear $f(Q)$ form, and the scalar field associated to the
non-trivial connection, are combined to give one canonical and one phantom
field in the minisuperspace Lagrangian. Hence, the combination of two
well-behaved fields appears effectively as a phantom field. This may open a
new way to handle the phantom fields, namely as effective realizations of
the connection-related canonical fields.

The plan of the work is the following: In Section \ref{fQgravity} we review
the basics of $f(Q)$ gravity and cosmology, presenting also the
minisuperspace approach. In Section \ref{secQuintom} we present the quintom
scenario, namely the simultaneous consideration of a canonical and a phantom
scalar field. Then in Section \ref{EquivSec} we show the dynamical
equivalence of the two scenarios. Finally, Section \ref{Conclusion} is
devoted to the Conclusions.

\section{$f(Q)$ gravity and cosmology}

\label{fQgravity}

In this section we intrdocue the  $f(Q)$ gravity and we apply it in a
cosmological framework.

\subsection{$f(Q)$ gravity}

We start by presenting the basics of symmetric teleparallel geometry. We
introduce a general affine connection as 
\begin{equation}
\Gamma _{\mu \nu }^{\alpha }=\hat{\Gamma}_{\mu \nu }^{\alpha }+K_{\,\,\mu
\nu }^{\alpha }+L_{\,\,\mu \nu }^{\alpha },
\end{equation}%
with $\hat{\Gamma}_{\mu \nu }^{\alpha }$ is the Levi-Civita connection, $%
K_{\,\,\mu \nu }^{\alpha }=\frac{1}{2}T_{\,\,\mu \nu }^{\alpha }+T_{(\mu
\,\,\,\nu )}^{\,\,\,\alpha }$ the contortion tensor related to the torsion
tensor $T_{\,\,\mu \nu }^{\alpha }$, and with $L_{\,\,\mu \nu }^{\alpha }=%
\frac{1}{2}Q_{\,\,\mu \nu }^{\alpha }-Q_{(\mu \,\,\,\nu )}^{\,\,\,\alpha }$
the disformation tensor which arises from the non-metricity tensor 
\begin{equation}
Q_{\alpha \mu \nu }\equiv \nabla _{\alpha }g_{\mu \nu },
\end{equation}%
with $g_{\mu \nu }$ the metric (throughout the manuscript Greek indices
denote coordinate space). 
In terms of the affine connection we can write the torsion, curvature and
nonmetricity tensors as 
\begin{eqnarray}
&&T^{\lambda }{}_{\mu \nu }\equiv \Gamma ^{\lambda }{}_{\mu \nu }-\Gamma
^{\lambda }{}_{\nu \mu }\,  \notag  \label{Tortnsor} \\
&&R^{\sigma }{}_{\rho \mu \nu }\equiv \partial _{\mu }\Gamma ^{\sigma
}{}_{\nu \rho }-\partial _{\nu }\Gamma ^{\sigma }{}_{\mu \rho }+\Gamma
^{\alpha }{}_{\nu \rho }\Gamma ^{\sigma }{}_{\mu \alpha }-\Gamma ^{\alpha
}{}_{\mu \rho }\Gamma ^{\sigma }{}_{\nu \alpha }\ \   \notag  \label{Rietsor}
\\
&&Q_{\rho \mu \nu }\equiv \nabla _{\rho }g_{\mu \nu }=\partial _{\rho
}g_{\mu \nu }-\Gamma ^{\beta }{}_{\rho \mu }g_{\beta \nu }-\Gamma ^{\beta
}{}_{\rho \nu }g_{\mu \beta }\,.  \label{NonMetrensor}
\end{eqnarray}

In general relativity one uses Riemannian geometry, in which case the
gravitational Lagrangian is just the Ricci scalar derived from contractions
of the Riemann tensor. Similarly, in the Teleparallel Equivalent of General
Relativity one uses the Weitzenb\"{o}ck geometry, in which case the
gravitational Lagrangian is just the torsion scalar derived from
contractions of the torsion tensor. Hence, one can obtain a third
description of gravity based on nonmetricity and the framework of symmetric
teleparallel geometry.

In particular, one can use as gravitational
Lagrangian the non-metricity scalar arising from contractions of the
non-metricity tensor, namely 
\begin{equation}
Q=-\frac{1}{4}Q_{\alpha \beta \gamma }Q^{\alpha \beta \gamma }+\frac{1}{2}%
Q_{\alpha \beta \gamma }Q^{\gamma \beta \alpha }+\frac{1}{4}Q_{\alpha
}Q^{\alpha }-\frac{1}{2}Q_{\alpha }\tilde{Q}^{\alpha }\,,
\label{NontyScalar}
\end{equation}%
with $Q_{\alpha }\equiv Q_{\alpha \ \mu }^{\ :\mu }\,,$ and $\tilde{Q}%
^{\alpha }\equiv Q_{\mu }^{\ :\mu \alpha }\,.$ Thus, similarly to the $f(R)$
and $f(\mathbb{T})$ extensions of the corresponding theories, one can extend the
Lagrangian of the symmetric teleparallel equivalent of general relativity to
an arbitrary function, obtaining $f(Q)$ gravity, with action \cite%
{BeltranJimenez:2017tkd}: 
\begin{equation}
S=\frac{1}{2\kappa }\int {\mathrm{d}}^{4}x\sqrt{-g}f(Q)+S_{m},
\label{fQaction}
\end{equation}%
with $\kappa =8\pi G$ the gravitational constant, and where we have also
included the matter sector action. Note that the Symmetric Teleparallel
Equivalent of General Relativity, with or without cosmological constant, is
obtained for linear function $f$, that is  $f_{,QQ}\left( Q\right) =0$, or
when $Q=$const.

Finally, performing variation yields the field equations of $f(Q)$ gravity
as \cite{BeltranJimenez:2019tme}: 
\begin{eqnarray}
&&
\!\!\!\!\!\!\!\!\!\!\!\!\!\!\!\!
\frac{2}{\sqrt{-g}}\nabla _{\lambda }(\sqrt{-g}f_{Q}P^{\lambda }{}_{\mu \nu
})-\frac{1}{2}fg_{\mu \nu }\nonumber\\
&&\ \ \ +f_{Q}(P_{\nu \rho \sigma }Q_{\mu 
}{}^{\rho
\sigma }-2P_{\rho \sigma \mu }Q^{\rho \sigma }{}_{\nu })=\kappa T_{\mu \nu
}^{m},  \label{fieldeq1}
\end{eqnarray}%
with $T_{\mu \nu }^{m}$ the matter energy-momentum tensor, $f_{Q}=\partial
f/\partial Q$, and where the superpotential $P^{\lambda }{}_{\mu \nu }$ is
written as {\small{
\begin{equation}
P^{\lambda }{}_{\mu \nu }=\frac{1}{4}\left( -2L^{\lambda }{}_{\mu \nu
}+Q^{\lambda }g_{\mu \nu }-\tilde{Q}^{\lambda }g_{\mu \nu }-\frac{1}{2}%
\delta _{\mu }^{\lambda }Q_{\nu }-\frac{1}{2}\delta _{\nu }^{\lambda }Q_{\mu
}\right) .  \label{Pdef}
\end{equation}}}
Additionally, variation of the action with respect to the general affine
connection leads the connection equation as 
\begin{equation}
\nabla _{\mu }\nabla _{\nu }(\sqrt{-g}f_{Q}P^{\nu \mu }{}_{\lambda })=0\,,
\label{ff1}
\end{equation}%
where we have assumed that the matter Lagrangian does not depend on the
connection.

\subsection{$f(Q)$ cosmology}

We proceed to the application of $f(Q)$ gravity at a cosmological framework.
We impose the flat Friedmann-Robertson-Walker (FRW) metric 
\begin{equation*}
ds^{2}=-N(t)^{2}dt^{2}+a\left( t\right) ^{2}\left( dr^{2}+r^{2}\mathrm{d}%
\theta ^{2}+r^{2}\sin ^{2}\theta \mathrm{d}\phi ^{2}\right) ,
\end{equation*}%
where $a(t)$ is the scale factor, and $N(t)$ is the lapse function. In this
case, the requirement the connection to be symmetric, flat and to inherit the
symmetries of the background spacetime, leads to three different families of
connections \cite{Hohmann:2020zre,Hohmann:2021ast,DAmbrosio:2021pnd, 
Paliathanasis:2023nkb,Paliathanasis:2015aos,Dimakis:2016mip,
Heisenberg:2022mbo, Shi:2023kvu}.
In particular, the 
nonzero connection
components for a flat FRW metric are 
\begin{align}
\Gamma _{~tt}^{t}& =K_{1}\,,~~\Gamma _{~rr}^{t}=K_{2}\,,~~\Gamma _{~\theta
\theta }^{t}=K_{2}r^{2}\,,  \notag  \label{Generalconnection} \\
\Gamma _{~tr}^{r}& =\Gamma _{~rt}^{r}=\Gamma _{~t\theta }^{\theta }=\Gamma
_{~\theta t}^{\theta }=\Gamma _{~t\phi }^{\phi }=\Gamma _{~\phi t}^{\phi
}=K_{3}\,,  \notag \\
\Gamma _{~r\theta }^{\theta }& =\Gamma _{~\theta r}^{\theta }=\Gamma
_{~r\phi }^{\phi }=\Gamma _{~\phi r}^{\phi }=\frac{1}{r}\,,~~\Gamma
_{~\theta \theta }^{r}=-r\,,  \notag \\
~~\Gamma _{~\phi \phi }^{r}& =-r\sin ^{2}\theta \,,~~\Gamma _{~\phi \phi
}^{t}=K_{2}r^{2}\sin ^{2}\theta \,,  \notag \\
\Gamma _{~\phi \theta }^{\phi }& =\Gamma _{~\theta \phi }^{\phi }=\cot
\theta \,,~~\Gamma _{~\phi \phi }^{\theta }=-\sin \theta \cos \theta \,,
\end{align}%
where $K_{1}(t),K_{2}(t),K_{3}(t)$ are functions of time, classified as

\begin{eqnarray}
&& \!\!\!\!\!\!\!\!\!\!\!\!\!\! \!\!\! \text{Connection 
I}\!:K_{1}=\gamma(t),K_{2}=0,K_{3}=0,
\\
&& \!\!\!\!\!\!\!\!\!\!\!\!\!\! \!\!\!\text{Connection II}\!:  
K_{1}=\frac{\dot{\gamma}(t)}{\gamma(t)}%
+\gamma(t),K_{2}=0,K_{3}=\gamma(t),\label{Connection2}
\\
 && \!\!\!\!\!\!\!\!\!\!\!\!\!\! \!\!\! \text{Connection III}\!:  
K_{1}=-\frac{\dot{\gamma}(t)}{\gamma(t)}%
,K_{2}=\gamma(t),K_{3}=0,
\end{eqnarray}
where $\gamma(t)$ a function of time.
Note that in the case of non-flat FRW geometry there is another connection,
namely Connection IV, which however recovers Connection III in the flat case 
 \cite{Hohmann:2020zre,Hohmann:2021ast,DAmbrosio:2021pnd, 
Paliathanasis:2023nkb,Paliathanasis:2015aos,Dimakis:2016mip,
Heisenberg:2022mbo, Shi:2023kvu}. 
Connection I has been widely studied in cosmological studies, however the
resulting background gravitational field equations        are
equivalent to that of $f(\mathbb{T}) $-teleparallel gravity as 
discussed
in \cite{Paliathanasis:2023nkb}.

In this work we consider the Connection II above. Thus, the nonmetricity
scalar is defined as 
\begin{equation}
Q\left( \Gamma \right) =-6H^{2}+\frac{3\gamma }{N}\left( 3H-\frac{\dot{N}}{%
N^{2}}\right) +\frac{3\dot{\gamma}}{N^{2}}.
\end{equation}%
Additionally, the general field equations (\ref{fieldeq1}) give rise to the
Friedmann equations as 
\begin{equation}
 3H^{2}f^{\prime }(Q)+\frac{1}{2}\left( f(Q)-Qf^{\prime }(Q)\right) +\frac{%
3\gamma \dot{Q}f^{\prime \prime }(Q)}{2N^{2}}=\kappa \rho _{m}, 
\end{equation}
\begin{eqnarray}
&&
\!\!\!\!\!\!\!\!\!\!\!\!\!\!\!\!\!
-\frac{2}{N}\frac{d}{dt}\left( f^{\prime }(Q)H\right) -3H^{2}f^{\prime }(Q)-%
\frac{1}{2}\left( f(Q)-Qf^{\prime }(Q)\right)\nonumber\\
&& \ \ \ \ \ \ \ \ \ \ \ \ \ \  
+\frac{3\gamma \dot{Q}%
f^{\prime \prime }(Q)}{2N^{2}}=\kappa p_{m},
\end{eqnarray}
where primes denote derivatives with respect to $Q$.

\subsection{Minisuperspace description}

Let us now formulate $f(Q)$ cosmology in the mini-superspace framework.
For simplicity in the following we set the gravitational constant $\kappa=1$ 
and we neglect the matter content. 
Consider the action in $f\left( Q\right) $ gravity, where we introduce the 
Lagrange
multiplier $\lambda ,~$such that%
\begin{equation}
S_{f\left( Q\right) }=\int \sqrt{-g}d^{4}x\left[ f\left( Q\right) -\lambda
\left( Q-\mathcal{Q}\right) \right] ,
\end{equation}%
where $Q\equiv \mathcal{Q}\left( N,\dot{N},a,\dot{a},\gamma ,\gamma \right) $
is the expression of the nonmetricity scalar. Variation with respect to $Q$,
gives $\lambda =f^{\prime }\left( Q\right) $. 

We replace in the action integral, and for the connection choice 
(\ref{Connection2}) we discussed in the previous subsection, it follows 
\begin{align*}
S_{f\left( Q\right) }& =\int dt\left[ Na^{3}\left( f\left( Q\right)
-f^{\prime }\left( Q\right) Q\right) -\frac{6}{N}a\dot{a}^{2}f^{\prime
}\left( Q\right) \right]  \\
& \ \ \ \ +\int dt\left[ 3\gamma f^{\prime }\left( Q\right) \left( 
\frac{3}{N}a^{2}%
\dot{a}-a^{3}\frac{\dot{N}}{N^{2}}\right) \right]  \\
& \ \ \ \ +\int dt\left[ \frac{3}{N}%
f^{\prime }\left( Q\right) a^{3}\dot{\gamma}\right] .
\end{align*}%
Integrating by parts the last term yields 
\begin{eqnarray}
&&
\!\!\!\!\!\!\!\!\!\!\!\!\!\!\!\!\!\!\!\!\!
\int dt\left[ \frac{3}{N}f^{\prime }\left( Q\right) a^{3}\dot{\gamma}\right]
= -\int dt\Big[ -\frac{3\dot{N}}{N^{2}}f^{\prime }\left( Q\right)
a^{3}\gamma\nonumber\\
&&
\ \ \ \ \ \ \ \ \ \ \   \ \ \ 
+\frac{9}{N}f^{\prime }\left( Q\right) 
a^{2}\dot{a}\gamma +\frac{%
3}{N}f^{\prime \prime }\left( Q\right) a^{3}\dot{Q}\gamma \Big]  \nonumber\\
&& \ \ \ \ \ \ \ \ \ \ \   \ \ \   +\text{surface terms}.
\end{eqnarray} 
Therefore, the point-like Lagrangian of the field equations is%
\begin{eqnarray}
&&
\!\!\!\!\!\!\!\!\!\!\!\!\!\!\!\!\!\!\!\!
L\left( \Gamma _{2}\right) =-\frac{3a\dot{a}^{2}f^{\prime }(Q)}{N}+\frac{N}{2%
}a^{3}\left[ f(Q)-Qf^{\prime }(Q)\right]\nonumber\\
&& \ \
-\frac{3a^{3}\dot{\psi}\dot{Q}%
f^{\prime \prime }(Q)}{2N},
\end{eqnarray}%
with the use of $\gamma =\dot{\psi}$.

We proceed by defining $\varphi =f^{\prime }\left( Q\right) $. Hence, we end
with the scalar-tensor Lagrangian 
\begin{equation}
L\left( \Gamma ^{B}\right) =-\frac{3}{N}\varphi a\dot{a}^{2}-\frac{3}{2N}%
a^{3}\dot{\varphi}\dot{\psi}-Na^{3}V\left( \varphi \right) ,  \label{lg2}
\end{equation}%
where $V(\varphi )$ is the effective scalar field potential, related to the $%
f\left( Q\right) $ function as%
\begin{equation}
V\left( f^{\prime }( Q) \right) =\frac{1}{2}\left[ Qf^{\prime}(Q)-f(Q)\right].
\end{equation}

\section{Quintom scenario}

\label{secQuintom}

In this section we will briefly review the quintom cosmology. Since we are 
interested in comparing with the minisuperspace description of $f(Q)$ 
cosmology,  we also  set the 
gravitational constant $\kappa=1$ 
and we neglect the matter content.

The quintom scenario
is based on the simultaneous consideration of two scalar fields: $\phi $,
which is canonical, and $\sigma $, which is phantom \cite{Feng:2004ad}. In
particular, the action is written as \cite{Cai:2009zp}: 
\begin{eqnarray}
&& \!\!\!\!\!\!\!\!\!\!\!\!\!\! \!\!\!\!\!\!\!\!\!\!\!\!\!\!\!
S =\int 
d^{4}x\sqrt{-g}\bigg[R-\frac{1}{2}g^{\mu \nu }\partial _{\mu }\phi
\partial _{\nu }\phi   \notag \\
&&
\ \ \ \ \ \ \ \,   
+\frac{1}{2}g^{\mu \nu }\partial _{\mu }\sigma \partial _{\nu }\sigma
+V\left( \phi ,\sigma \right)\bigg],  \label{actionquint}
\end{eqnarray}%
with  $V\left( \phi ,\sigma \right)$ the potential (one can split it in two 
separate potentials, such as $V\left( \phi ,\sigma \right)=V_{\phi }(\phi 
)+V_{\sigma }(\sigma )$, however this is not necessary. 

In a flat FRW geometry, variation of the above action with respect to the
metric gives the Friedmann equations \cite{Cai:2009zp}: 
\begin{equation}
H^{2}=\frac{1}{3}\Big[\frac{1}{2N^{2}}\left(\dot{\phi}^{2} 
- \dot{\sigma}^{2}\right)+V\left( \phi ,\sigma \right) \Big],  
\label{FR1quintom}
\end{equation}%
\begin{equation}
\dot{H}=-\frac{1}{2}\Big[\frac{1}{N^{2}}\left(\dot{\phi}^{2} 
- \dot{\sigma}^{2}\right)
\Big].  \label{FR2quintom}
\end{equation}
Additionally, variation of the action with respect to the scalar fields gives
the Klein-Gordon equations as 
\begin{eqnarray}
&&\frac{1}{N}\left( \frac{\dot{\phi}}{N}\right) ^{\cdot }+\frac{3}{N}H\dot{%
\phi}+\frac{\partial V\left( \phi ,\sigma \right)}{\partial \phi }=0  
\label{sddot} \\
&&\frac{1}{N}\left( \frac{\dot{\sigma}}{N}\right) ^{\cdot }+\frac{3}{N}H\dot{%
\sigma}-\frac{\partial V\left( \phi ,\sigma \right)}{\partial \sigma }=0.
\end{eqnarray}%

In quintom cosmology, the effective dark energy is attributed to the
canonical and phantom fields, and thus its energy density and pressure are
given by 
\begin{eqnarray}
&&\rho _{DE}\equiv \rho _{\phi }+\rho _{\sigma 
}=\frac{1}{2N^{2}}\left(\dot{\phi}^{2} 
- \dot{\sigma}^{2}\right)+V\left( \phi ,\sigma \right) \label{rhoDE} \\
&&p_{DE}\equiv p_{\phi }+p_{\sigma }=\frac{1}{2N^{2}}\left(\dot{\phi}^{2} 
- \dot{\sigma}^{2}\right)-V\left( \phi ,\sigma \right),  \label{pDE}
\end{eqnarray}%
   while its equation-of-state parameter is 
\begin{equation}
w_{DE}\equiv \frac{p_{DE}}{\rho _{DE}}= \frac{\left( 
\dot{\phi}^{2}-\dot{\sigma}%
^{2}\right) -2N^{2}V\left( \phi ,\sigma \right) }{\left( \dot{\phi}^{2}-\dot{%
\sigma}^{2}\right) +2N^{2}V\left( \phi ,\sigma \right) }.
\end{equation}%

\section{Dynamical equivalence of $f(Q)$ cosmology with Quintom scenario}

\label{EquivSec}

In this section we will show the dynamical equivalence of $f(Q)$ cosmology,
in the minisuperspace framework, with the quintom scenario. Considering that
in (\ref{lg2}) the lapse function is $N=\bar{N}\varphi $, it follows 
\begin{equation}
L\left( \Gamma ^{B}\right) =-\frac{3}{\bar{N}}a\dot{a}^{2}-\frac{3}{2\bar{N}}%
a^{3}\frac{\dot{\varphi}}{\varphi }\dot{\psi}-\bar{N}a^{3}\left( \varphi
V\left( \varphi \right) \right) .
\end{equation}%
Defining $\Phi =\ln \varphi $ we acquire 
\begin{equation}
L\left( \Gamma ^{B}\right) =-\frac{3}{\bar{N}}a\dot{a}^{2}-\frac{3}{2\bar{N}}%
a^{3}\dot{\Phi}\dot{\psi}-\bar{N}a^{3}\left( e^{\Phi }V\left( \Phi \right)
\right) .
\end{equation}%
Finally, introducing 
\begin{eqnarray}
\Phi  &=&\frac{1}{\sqrt{3}}\left( \phi +\sigma \right)  \\
\psi  &=&\frac{1}{\sqrt{3}}\left( \sigma -\phi \right) ,
\end{eqnarray}%
equivalently we have 
\begin{equation}
L\left( \Gamma ^{B}\right) =-\frac{3}{\bar{N}}a\dot{a}^{2}+\frac{1}{2\bar{N}}%
a^{3}\left( \dot{\phi}^{2}-\dot{\sigma}^{2}\right) -\bar{N}a^{3}\hat{V}%
\left( \phi +\sigma \right) ,
\end{equation}%
where $\hat{V}\left( \phi +\sigma \right) =e^{\left( \phi +\sigma \right)
}V\left( \phi +\sigma \right) $. In this case, the first Friedmann equation
is written as 
\begin{equation}
3\bar{H}^{2}=\frac{1}{2\bar{N}^{2}}\left( \dot{\phi}^{2}-\dot{\sigma}%
^{2}\right) +\hat{V}\left( \phi +\sigma \right) ,
\end{equation}%
with $\bar{H}=\frac{1}{\bar{N}}\frac{\dot{a}}{a}$, and where we have defined 
$H=\frac{1}{\phi +\sigma }\bar{H}$.
Consequently, we can define the    effective dark energy sector with energy 
density and pressure
\begin{eqnarray}
&&
\!\!\!\!\!\!\!\!\!\!\!\!
\rho _{DE}^{Q}\equiv \rho _{\phi }+\rho _{\sigma 
}=\frac{1}{2\bar{N}^{2}}\left( \dot{\phi}^{2}-\dot{%
\sigma}^{2}\right) +\hat{V}\left( \phi +\sigma \right) \label{rhoDE22} \\
&&
\!\!\!\!\!\!\!\!\!\!\!\!
p_{DE}^{Q}\equiv p_{\phi }+p_{\sigma }=\frac{1}{2\bar{N}^{2}}\left( 
\dot{\phi}^{2}-\dot{\sigma}%
^{2}\right) -\hat{V}\left( \phi +\sigma \right) ,  \label{pDE22}
\end{eqnarray}
such that the equation-of-state parameter is defined as%
\begin{equation}
w_{DE}\equiv \frac{p_{DE}}{\rho _{DE}}= \frac{\left( 
\dot{\phi}^{2}-\dot{\sigma}%
^{2}\right) -2N^{2}V\left( \phi ,\sigma \right) }{\left( \dot{\phi}^{2}-\dot{%
\sigma}^{2}\right) +2N^{2}V\left( \phi ,\sigma \right) }.
\end{equation}%

As we see, $f(Q)$ cosmology with the second connection gives rise to a
two-field model. In particular, one field comes from the non-linear $f(Q)$
form, and the other one arises from the extra connection field $\gamma $,
while their combination yields the final fields, $\phi $ and $\sigma $, that
appear in the Friedmann equation. Interestingly enough, we see that one
field appears in the standard canonical way, while the other appears with
opposite kinetic energy, ans thus exhibiting effectively a phantom behavior.
Thus, $f(Q)$ cosmology with the second connection is dynamically equivalent
with a quintom-like model. We use the term quintom-like and not quintom, since 
we have obtained a nonminimal coupling through $H=\frac{1}{\phi +\sigma }\bar{H}
$, i.e. we obtain an effective gravitational constant proportional to $1/({\phi 
+\sigma })^2$. However, dynamically any trajectory of the basic quintom model 
is equivalent to the $f\left( Q\right) $ cosmology. The observable physics may 
be different, except in the asymptotic limit where $\phi +\sigma $ is constant, 
that is  $Q$ becomes constant and the limit of General Relativity is recovered.

We mention here that one can obtain a full equivalence between $f(Q)$ 
cosmology and the quintom scenario, if he extends to nonminimal coupled 
quintom. In this case, using terms of the form $ f(\phi) R$ and  
$f(\sigma)  R$ gives rise to Friedmann equations with effective 
gravitational constant, where $H^2$ is multiplied by functions of $\phi$ and 
$\sigma$  \cite{Setare:2008pc}. Additionally, one can straightforwardly 
consider the matter sector too, without or with nonminimal couplings, or 
interactions with the scalar fields.

Nevertheless, the important result that we desire to point out in this 
first  paper on the subject, is that we have provided a
way to obtain a phantom field from first principles, namely from a
non-trivial connection. Indeed, in the rich literature of phantom and
quintom fields and their applications, the phantom field is introduced in the 
action ad hoc, and thus one typically faces all the known problems that
the phantom fields could have at the quantum level, since the energy is
unbounded from below. However, in this work we showed that what appears
effectively as a phantom field, is the combination of the well-behaved $f(Q)$
scalar field together with the well-behaved connection field $\gamma $ of
the non-trivial connection. In summary, this may open a new way to handle
the phantom fields, namely as effective realizations of the
connection-related canonical fields.

Regarding the dynamical degrees of freedom of the theory, we mention that   
in the  
connections II and III  the nonmetricity scalar has a
boundary term contribution which depends on  $\gamma $. Due to this boundary 
term, the nonmetricity has second-order
derivative terms of the scale factor  and of the derivative of $\gamma$. 
Consequently, a nonlinear function $f\left( Q\right) $  leads to an
analogy of a fourth-order theory,
with the
additional scalar $\dot{\psi}=\gamma $. Hence, by using a Lagrange
multiplier we were able to re-write the fourth-order theory as a second-order 
one, by
increasing the dependent variables (introducing a second scalar field), and due 
to the structure of the theory    the combination of the two
  fields appears effectively as a phantom field.
Nevertheless, for the connection I, the non-metricity does not have any
boundary term contribution, and it depends only on first-order derivatives
of the scale factor, leading finally to a second-order theory of gravity.

\section{Conclusion}
\label{Conclusion}

We showed that $f(Q)$ cosmology  is dynamical equivalent with a quintom-like 
model. $f(Q)$ gravity is a modified theory of gravity that uses the 
non-metricity tensor as the fundamental geometric quantity which describes the  
gravitational field, and its cosmological applications are very rich and have 
been widely investigated. On the other hand, the quintom scenario is based on 
the simultaneous consideration of one canonical and one phantom scalar field, 
and its cosmological phenomenology is very interesting since it can naturally 
realize the phantom-divide crossing for the dark energy sector. Nevertheless,  
the introduction of the phantom field in phantom and quintom cosmology is ad 
hoc, and the fact that the energy of the phantom field is unbounded from below 
makes it potentially unstable at the quantum level.
 
We formulated $f(Q)$ cosmology in the minisuperspace approach, 
considering also the non-trivial connection, namely Connection II of the 
literature. In this case the scalar field arising
from the non-linear $f(Q)$ form, and the scalar field associated to the
non-trivial connection, are combined to give one canonical and one phantom
field in the minisuperspace Lagrangian. In other words, the combination of two
well-behaved fields appears effectively as a phantom field. The resulting 
Lagrangian and Friedmann equations are dynamically equivalent with a 
quintom-like scenario.

In summary, although in the standard approach to phantom and quintom scenario, 
the phantom field is added by hand, which then may raise issues concerning the 
stability, in the present analysis we showed that the phantom fields can arise 
as effective realizations of the connection-related canonical fields, and thus 
the quintom scenario is just a particular case of $f(Q)$ cosmology. This 
correspondence may open a new avenue to handle phantom fields and their 
applications in a cosmological framework.

 We close this work with the following comments.  $f\left( Q\right) $ 
gravity is known to exhibit some pathologies, investigated in detail in
\cite{Gomes:2023tur}, where it was found that strong
coupling and ghosts appear. However, in  \cite{Guzman:2024cwa} it was found that 
connection II offers possibilities for a realistic cosmological description,
assuming a specific set of initial conditions. The scalar-field formalism, that 
we presented in this work, can be used to investigate further this subject. 
Additionally, it is worth to investigate the conformal transformation of the 
theory. Finally, one powerful approach that allows to study  the existence of 
pathologies in a theory is the  projective symmetry 
\cite{BeltranJimenez:2019acz,BeltranJimenez:2020sqf}, since projective 
invariance is a sufficient condition for the absence of  ghost-like 
instabilities. Hence, examining the projective symmetry of $f(Q)$ theory with   
connection II is both interesting and necessary, and can reveal important 
information about the structure of the theory. Since such a full analysis is out 
of the scopes of this work, it will be performed in a future project.

\end{document}